\documentstyle[12pt,epsf]{article}

\newcommand{\slsh}[1]{\, {\not {\! #1}}}
\newcommand{\oabs}[1]{|\vec{#1} \makebox[0.1em]{} |}
\newcommand{\oabsq}[1]{{|\vec{#1} \makebox[0.1em]{} |}^2}
\begin{document}

\begin{center}
{\bf CONSISTENT EFFECTIVE DESCRIPTION OF NUCLEONIC RESONANCES IN AN UNITARY
RELATIVISTIC FIELD--THEORETIC WAY}
\footnote{FAU-TP3-98/11; invited talk given at XIV. Int.\ Sem.\ on High En.\
Phys.\ Probl.,
 17.-22.8.1998, Dubna}

\vskip 5mm
F. Kleefeld

\vskip 5mm

{\small
{\it
Institute for Theoretical Physics III, University of Erlangen-N\"urnberg,
}
\\
{\it
Staudtstr.\ 7, 91058 Erlangen, Germany
}
\\
{\it
E-mail: kleefeld@theorie3.physik.uni-erlangen.de
}}
\end{center}

\vskip 5mm

\begin{center}
\begin{minipage}{150mm}
\centerline{\bf Abstract}
High energy strong interaction physics is successfully described by the local
renormalizable gauge theory called Quantum--Chromo--Dynamics (QCD) with quarks and
gluons as ``elementary''
degrees of freedom, while intermediate
energy strong interaction physics shows up to be determined by a non--local,
non--renormalizable effective field theory (EFT) of ``effective'' degrees of 
freedom like
mesons, ground state baryons and resonances. The connection between high and
intermediate physics is established by a change of basis (``bosonisation'') 
from the infinite
Fock--state basis of quarks and gluons to the infinite Fock--state basis of the
``effective'' degrees of freedom.
The infinite number of counter terms in the Lagrangian of such an 
non--renormalizable EFT is 
replaced by a tree--level Lagrangian containing a finite number of interaction
terms dressed by non--local vertex--functions commonly called formfactors 
(containing cutoffs) generating the dynamics of an infinity of
interaction diagrams in an EFT. Furthermore low and intermediate energy
physics successfully is described by the use of resonance propagators, i.e.
resonances are treated like ``degrees of freedom'', which are seen in the 
experiment and behave like particles with complex mass which is usually not 
compatible with the idea of unitarity.\\
In analogy to the role of vertex-functions in non--renormalizable theories and with
respect to the infinite dimension of the effective Fock-state basis I 
present a ``toy-model'' in which fermionic and bosonic resonances are considered
to be
``particles'', i.e.\ they consistently are described by (anti-)commuting
effective field-operators (containing dynamics of infinitely many
quark-gluon or meson-nucleon diagrams) which are comfortably treated by
Wick's Theorem in a covariant framework and obey unitarity. Non--trivial implications to
couplings of non--local interactions are shown.
\end{minipage}
\end{center}
\vskip 10mm
\section{Introduction}
\begin{figure}[tp]
\epsfxsize=  16.0cm
\epsfysize=  9.5cm
\centerline{\epsffile{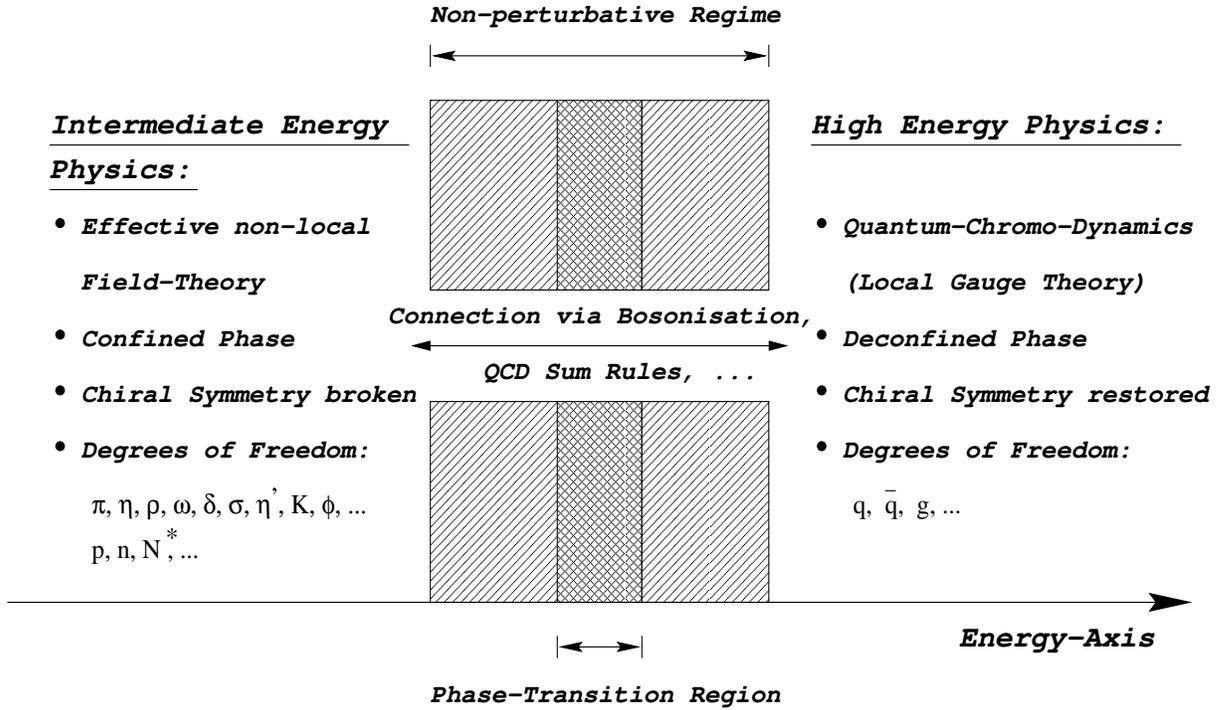}}
\caption{Confrontation of intermediate and high energy strong interaction physics.} \label{figg1}
\end{figure}

Why does a intermediate energy theorist hold a talk on a high energy physics and
QCD conference? The answer of this question is obvious: The high and
the intermediate energy approaches to strong interaction physics are treating 
the same problem, i.e. revealing the physical nature of the phase transition 
region (see Fig.\ \ref{figg1}), from two different directions. Both approaches are connected on
fundamental theoretical grounds, use similar techniques and suffer more or less
similar problems in the non--perturbative regime. The solution to the problem of 
the phase transition consists of a consistent combination of the theoretical 
framework of both sides forming one fundamental framework describing the 
theory of strong interaction. Hopeful steps in this direction based on chiral
symmetry constraints and analyticity properties lead e.g.\ to Chiral Perturbation
Theory \cite{gas1,gas2,ber1} and the QCD Sum Rule method \cite{shi1,iof1,nam1}.\\
Using the constraint of unitarity I want to study the property of self--energies
and vertex--functions (``formfactors'') in nonlocal effective (bosonised)
intermediate energy field theories. The reason is simple: The experimental
situation in intermediate energy proton--proton and proton--nucleus colliders 
improved a lot. In cooled synchrotrons (e.g. at COSY, WASA, $\ldots$) the
experimentalists produce high precision datas for exclusive meson production
processes at threshold like $pp\rightarrow pp\,\pi^0$, $pp\rightarrow pp\,\eta$,
$pp\rightarrow p\Lambda K$, $pp\rightarrow pp\,\phi$, $pp\rightarrow
pp\,\pi\pi$, $pd\rightarrow {}^3H\!e\,\eta$ , $\ldots\,$. The high energy and momentum transfers involved in
these reactions excite all kinds of effective degrees of freedom like hyperons, 
resonances, $\ldots\,$. The physics involved for heavy meson production is so
short ranged that intermediate energy theorists have to leave common grounds and
describe the short ranged processes by high energy approaches in the
non--perturbative regime which involve quarks and gluons. The formfactors
mentioned contain all kinds of thinkable short range physics which have to be
revealed to get an understanding what's going on in the non--perturbative range
between intermediate and high energy physics. Furthermore the comparison of
experimental datas and theoretical calculations show a high sensitivity of
different isospin--channels to interference effects between various subprocesses
leading to the production of the considered mesons. Interference effects are
connected to the imaginary parts of production amplitudes which are due to
complex self--energies of intermediate resonances, loop contributions and --- as I will show ---
complex vertex functions at the interaction vertices.
The contradictions arrising with respect to time reversal invariance and
unitarity of  
theory can be resolved by a systematic introduction of effective degrees of
freedom and the application of certain constraints to vertex--functions and
self--energies due to unitarity.\\ 
As a final remark I want to mention that some of the present interests of
intermediate energy theorists should be very common to high energy physicists:
e.g. the question on the strangeness and spin content of the proton and its
excitations, the investigation of
Zweig-rule violations, the nature of non--abelian non--linear field-theories,
$\ldots\,$.  
\section{A simple classification scheme for resonances}
From Quantum Mechanics we know that close to a resonance $\cot\delta_\ell (E)$ is
variing rapidly as a function of the energy $E$, i.e. we can expand  
$\cot\delta_\ell (E)$ at the resonance energy $E_R$:
\begin{equation} \cot\delta_\ell (E) \quad = \quad  
\underbrace{\cot\delta_\ell (E_R)}_{= 0} \; + \; 
\frac{\partial\cot\delta_\ell (E)}{\partial E}\Bigg|_{E=E_R} (E-E_R) \; + \; \ldots 
\end{equation}
Using this expansion it simple to see that the partial scattering aplitude
$f_\ell (E)$ develops a Breit--Wigner shape (with a partial width $\Gamma_\ell$):
\begin{equation} f_\ell (E) = \frac{1}{k (\cot\delta_\ell - i )} 
\approx - \; \frac{1}{k_R} \; \frac{\Gamma_\ell / 2}{E-E_R + i \Gamma_\ell /2} \quad
\mbox{with} \quad \frac{2}{\Gamma_\ell} := - 
\frac{\partial\cot\delta_\ell (E)}{\partial E}\Bigg|_{E=E_R}  
\label{ggam1}\end{equation}
It is now important to mention that --- although the T--matrix develops an
imaginary part --- in this class of resonances there is no inelasticity present, i.e.
{\em unitarity is still fulfilled, if the partial phaseshifts $\delta_\ell$ are 
real}.\\
In a second class of resonances absorptivity, i.e. inelasicity is present,
sometimes desired. In such a case the whole system looses probability. 
In the well known ``Wigner--Weisskopf approximation'' the
decay of particles is described by a non--Hermitian Hamilton--operator. These
decaying particles have (like resonances) a finite decay width $\Gamma$ which determine the
non--diagonal elements of the Hamilton--operator. The reason for the
inelasticity in this kind of approaches is, that parts of the Hilbert--space
have been removed from the problem, i.e. the Hilbert--space/Fock--space is
incomplete. Examples for this approach are the optical potential method and
the description of $K^0$-$\bar{K}^0$--oszillations.\\
\section{Self--energies and vertex--functions
in renormalizable and non-renormalizable field theories}
Looking at equation (\ref{ggam1}) one could assume that the self--energy or
resonance width is constant. That this is not the general case can be seen from
the normalizable local gauge theory QED. Here the electron propagator 
$G^{\bar{e}e}(-p,p)= i (\!\!\slsh{p}-m_e+\Sigma^{\bar{e}e}(p))^{-1}$ develops by the
renormalization procedure a momentum dependent self--energy 
$\Sigma^{\bar{e}e}(p)$. To one loop one obtains:
\begin{equation} 
\Sigma^{\bar{e}e}(p) = i\,e^2 \,\int\frac{d^4q}{(2\pi)^4} 
\frac{\gamma_\alpha (\slsh{q}+m_e) \,\gamma^\alpha}{(q^2-m^2_e)((q-p)^2-\lambda^2)}
\end{equation}   
In the same way the bare electron--photon--vertex picks up a momentum
dependence, i.e. $i\,e\,\gamma^\mu \rightarrow i\,e\,\gamma^\mu + i\,e\,\Lambda^\mu
(p^\prime , p)$. Examples for self--energies developing momentum
dependent imaginary parts are quasi--particle excitations in finite density
field--theories and the famous ``Landau--damping''. Of course, the contact to a
heat bath or a finite medium locally violates unitarity, i.e. is inelastic. But how about
resonances in effective field theories propagating in the vacuum? If resonant
effective degrees of freedom are present in an effective  
field theory of strong interaction, the corresponding effective Lagrangian should
not contain inelasticities, i.e. it should be Hermitian, as it should be derivable
by an unitary transformation from the Hermiatian Lagrangian of QCD. In such a
transformation the gluons have to be
integrated out from the generating functional (which is not possible at
present). After ``Fierz-ing'' properly the so obtained
multi-quark-Lagrangian and introducing source-terms for mesons {\em and baryons}
with all kinds of quantum numbers, the quark-fields have to be integrated out,
to obtain an non--local effective action of all the mesonic and baryonic
sources. Finally the infinity of interaction terms in the non-local
Lagrangian obtained
have to be replaced by a tree-level Lagrangian containing {\em complex} momentum dependent
vertex--functions and self--energies. The
following toy model will show that resonance sources have to appear {\em pairwise} due
to unitarity.
\section{The ``Unitary Effective Resonance Model''}
\subsection{Field theoretic effective model for one fermionic resonance}
At this point I only roughly sketch a toy model developed for the consistent
description of effective resonant degrees of freedom in intermediate energy
strong interaction physics. For a more complete view I refer to references
\cite{kle1}\cite{kle2}.\\
It is very common in intermediate energy physics to describe the propagation of
a fermionic resonance with the complex ``mass'' $M=m_\ast-i\,\Gamma /2$ 
($m_\ast$ is the real part of the mass $M$, $\Gamma$ is the resonance width) 
by the introduction of a propagator of the form $(\!\!\slsh{p}-M)^{-1}$. That
this simplistic picture is not well defined can be seen, if one ``naively''
tries to write down a free Lagrangian for a fermionic resonance field $\Psi (x)$ in the
following way:
\begin{equation} {\cal L}_0 (x) \quad \stackrel{?}{=} \quad
\bar{\Psi} (x) \;
\bigg( \frac{1}{2} \; (i \! \slsh{\partial} 
                  - i \!\! \stackrel{\leftarrow}{\slsh{\partial}} ) 
                  - m_\ast + \frac{i}{2} \, \Gamma
\bigg) \; \Psi (x) \quad 
\end{equation}  
The action $S_0=\int\!d^4x\,{\cal L}_0 (x)$ related to this Lagrangian is not 
Hermitian. But Hermiticity of the action is the minimum requirement for
a CPT-invariant and unitary theory describing strong interaction only.
Additionally, if one tries to ``quantize'' such an effective resonance field,
i.e. introduce field operators, anticommutation relations, $\ldots$ one immediately runs into
inconsistencies.
To avoid all mentioned troubles one 
is forced to introduce {\em two} distinct field operators
$\bar{\Psi}_L (x)$ und $\Psi_R (x)$ (``left'' and ``right'' eigen-fields) and
their complex conjugates, to describe {\em one} resonance degree of freedom in an
effective field theoretic way by the following free Lagrangian density:
\begin{eqnarray} {\cal L}_0 (x) & = &
\alpha \; \bar{\Psi}_L (x) \;
\bigg( \frac{1}{2} \; (i \! \slsh{\partial} 
                  - i \!\! \stackrel{\leftarrow}{\slsh{\partial}} ) 
                  - m_\ast + \frac{i}{2} \, \Gamma
\bigg) \; \Psi_R (x) \quad + \nonumber \\
 & & \nonumber \\
 & + &
\alpha^\ast \; \bar{\Psi}_R (x) \;
\bigg( \frac{1}{2} \; (\underbrace{i \! \slsh{\partial} 
                  - i \!\! \stackrel{\leftarrow}{\slsh{\partial}}
                  }_{\displaystyle =: i \!\!
                  \stackrel{\;\,\leftrightarrow}{\slsh{\partial}}} ) 
                  - m_\ast - \frac{i}{2} \, \underbrace{\gamma_0 \, \Gamma^+
                  \, \gamma_0}_{\displaystyle \stackrel{!}{=} \Gamma}
\bigg) \; \Psi_L (x) \nonumber   
\end{eqnarray}
\newpage
\noindent ($\alpha$ is an arbitrary complex constant chosen to be 1). The classical
Euler--Lagrange equations with respect to variation of the action by
$\Psi^+_L (x)$,
$\Psi^+_R (x)$, $\Psi_L (x)$ and $\Psi_R (x)$ are the following generalized
"Dirac-equations":
\[
\begin{array}{rcr}
 {(i \! \slsh{\partial} - M ) \; \Psi_R (x) \quad = \quad 0}
 & \qquad \qquad &
 {\bar{\Psi}_R (x) (- i \!\! \stackrel{\leftarrow}{\slsh{\partial}} - M^\ast )
 \quad = \quad 0} \\
 {(i \! \slsh{\partial} - M^\ast ) \; \Psi_L (x) \quad = \quad 0} 
 & \qquad \qquad &
 {\bar{\Psi}_L (x) (- i \!\! \stackrel{\leftarrow}{\slsh{\partial}} - M ) 
 \quad = \quad 0} 
\end{array}
\]
\begin{equation} \end{equation}
These classical equations of motion can be solved by a simple
Laplace--transformation. The corresponding transformation for the
field--operators is:
\begin{eqnarray}
\Psi_R (x) & = & \sum\limits_{s} \, \int \! \!
\frac{d^3 k \; \sqrt{2 M}}{\sqrt{(2\pi )^3 \; 2 \omega_R \, (\oabs{k} )}}
\Big[ 
 u_R \, (\vec{k} , s ) \; b_R \, (\vec{k} , s ) \; e^{\displaystyle - \, i k_R
 x} +
 v_R \, (\vec{k} , s ) \; d^+_R \, (\vec{k} , s ) \; e^{\displaystyle i k_R x}
\Big] \nonumber \\
 & & \nonumber \\
\Psi_L (x) & = & \sum\limits_{s} \, \int \! \!
\frac{d^3 k \; \sqrt{2 M^\ast}}{\sqrt{(2\pi )^3 \; 2 \omega_L \, (\oabs{k} )}}
\Big[ 
 u_L \, (\vec{k} , s ) \; b_L \, (\vec{k} , s ) \; e^{\displaystyle - \, i k_L
 x} +
 v_L \, (\vec{k} , s ) \; d^+_L \, (\vec{k} , s ) \; e^{\displaystyle i k_L x}
\Big] \nonumber \\
 & & \nonumber \\
\bar{\Psi}_R (x) & = & \sum\limits_{s} \, \int \! \!
\frac{d^3 k \; \sqrt{2 M^\ast}}{\sqrt{(2\pi )^3 \; 2 \omega_L \, (\oabs{k} )}}
\Big[ 
 \bar{v}_R \, (\vec{k} , s ) \; d_R \, (\vec{k} , s ) \; e^{\displaystyle - \,
 i k_L x} +
 \bar{u}_R \, (\vec{k} , s ) \; b^+_R \, (\vec{k} , s ) \; e^{\displaystyle i
 k_L x}
\Big] \nonumber \\
 & & \nonumber \\
\bar{\Psi}_L (x) & = & \sum\limits_{s} \, \int \! \!
\frac{d^3 k \; \sqrt{2 M}}{\sqrt{(2\pi )^3 \; 2 \omega_R \, (\oabs{k} )}}
\Big[ 
 \bar{v}_L \, (\vec{k} , s ) \; d_L \, (\vec{k} , s ) \; e^{\displaystyle - \,
 i k_R x} +
 \bar{u}_L \, (\vec{k} , s ) \; b^+_L \, (\vec{k} , s ) \; e^{\displaystyle i
 k_R x}
\Big] \nonumber 
\end{eqnarray}
\begin{equation} \end{equation}
with $\omega_R \, ( \oabs{k} ) := \sqrt{\oabsq{k} + M^2}\;$, 
$ \omega_L \, ( \oabs{k} ) := \sqrt{\oabsq{k} + M^{\ast \, 2}} \;$,
$k^{\,\mu}_R :=
(\omega_R \, ( \oabs{k} ) , \, \vec{k} \,\, )$ and $k^{\,\mu}_L := (\omega_L \,
( \oabs{k} ) , \, \vec{k} \,\, )$.
The generalized ``Dirac--spinors'' solving the corresponding momentum space
``Dirac--equations'':
\begin{eqnarray} 
 (- \, \slsh{k}_R + M ) \; u_R \, (\vec{k} , s) \quad = \quad 0 & \quad, \quad
 & 
 \bar{u}_R \, (\vec{k} , s) \; (- \, \slsh{k}_L + M^\ast ) \quad = \quad 0
 \nonumber \\
 (- \, \slsh{k}_L + M^\ast ) \; u_L \, (\vec{k} , s) \quad = \quad 0 & \quad
, \quad & 
 \bar{u}_L \, (\vec{k} , s) \; (- \, \slsh{k}_R + M ) \quad = \quad 0 \nonumber
 \\
 (\slsh{k}_R + M ) \; v_R \, (\vec{k} , s) \quad = \quad 0 & \quad ,\quad & 
 \bar{v}_R \, (\vec{k} , s) \; (\slsh{k}_L + M^\ast ) \quad = \quad 0 \nonumber
 \\
 (\slsh{k}_L + M^\ast ) \; v_L \, (\vec{k} , s) \quad = \quad 0 & \quad ,\quad
 & 
 \bar{v}_L \, (\vec{k} , s) \; (\slsh{k}_R + M ) \quad = \quad 0   
\end{eqnarray}
are found to be:
\begin{eqnarray}
 u_R \, (\vec{k} , s) \; := & \displaystyle 
 \frac{\slsh{k}_R + M}{\sqrt{2 M (M + \omega_R \, ( \oabs{k} ) )}} \; u_R \,
 (\vec{0} , s) 
& = \;
\sqrt{\frac{M + \omega_R \, ( \oabs{k} )}{2 M}} 
\left( \begin{array}{r} {\displaystyle \varphi_s} \\
 {\displaystyle \frac{\vec{\sigma} \cdot \vec{k}}{M + \omega_R \, ( \oabs{k} )}
 \; \; \varphi_s}
\end{array} \right) \nonumber \\
 & & \nonumber \\
 u_L \, (\vec{k} , s) \; := & \displaystyle 
 \frac{\slsh{k}_L + M^\ast}{\sqrt{2 M^\ast (M^\ast + \omega_L \, ( \oabs{k} )
 )}} \; u_L \, (\vec{0} , s) 
& = \;
\sqrt{\frac{M^\ast + \omega_L \, ( \oabs{k} )}{2 M^\ast}} 
\left( \begin{array}{r} {\displaystyle \varphi_s} \\
 {\displaystyle \frac{\vec{\sigma} \cdot \vec{k}}{M^\ast + \omega_L \, (
 \oabs{k} )} \; \; \varphi_s}
\end{array} \right) \nonumber \\
 & & \nonumber \\
 v_R \, (\vec{k} , s) \; := & \displaystyle 
 \frac{- \, \slsh{k}_R + M}{\sqrt{2 M (M + \omega_R \, ( \oabs{k} ) )}} \; v_R
 \, (\vec{0} , s) 
& = \;
\sqrt{\frac{M + \omega_R \, ( \oabs{k} )}{2 M}} 
\left( \begin{array}{r} 
 {\displaystyle \frac{\vec{\sigma} \cdot \vec{k}}{M + \omega_R \, ( \oabs{k} )}
 \; \; \chi_s} \\
{\displaystyle \chi_s} 
\end{array} \right) \nonumber \\
 & & \nonumber \\
 v_L \, (\vec{k} , s) \; := & \displaystyle 
 \frac{- \, \slsh{k}_L + M^\ast}{\sqrt{2 M^\ast (M^\ast + \omega_L \, ( \oabs{k}
 ) )}} \; v_L \,
 (\vec{0} , s) 
& = \;
\sqrt{\frac{M^\ast + \omega_L \, ( \oabs{k} )}{2 M^\ast}} 
\left( \begin{array}{r} 
 {\displaystyle \frac{\vec{\sigma} \cdot \vec{k}}{M^\ast + \omega_L \, (
 \oabs{k} )} \; \;
 \chi_s} \\
{\displaystyle \chi_s} 
\end{array} \right) \nonumber 
\end{eqnarray}
\begin{equation} \end{equation}
Canonical quantization in configuration space yields 
the following non--vanishing equal--time anticommutation relations:
\[
{\{ \, \Psi_{R,\, \sigma} (\vec{x},t) , \Psi^+_{L,\, \tau} (\vec{y},t) \,
\} 
 \quad = \quad \delta^{\, 3} (\vec{x} - \vec{y}) \; \delta_{\sigma \tau}} 
\qquad \mbox{\& Hermitian conjugate}
\]
In momentum space consistency implies for the non--vanishing anticommutators:
\begin{eqnarray} & &
{\{ \, b_R \, (\vec{k} , s ) ,  \, b^+_L \, (\vec{k}^\prime , s^\prime )
\} 
 \; = \; \delta^{\, 3} (\vec{k} - \vec{k}^\prime ) \;
 \delta_{ss^\prime}} 
 \quad , \quad
{\{ \, d_L \, (\vec{k} , s ) ,  \, d^+_R \, (\vec{k}^\prime , s^\prime )
\} 
 \; = \; \delta^{\, 3} (\vec{k} - \vec{k}^\prime ) \;
 \delta_{ss^\prime}} \nonumber \\
 & & \qquad \qquad \qquad \qquad \qquad \qquad \qquad \mbox{\& Hermitian conjugates} 
\end{eqnarray}
It is straight forward to construct the ``Feynman--propagators'' by:
\begin{eqnarray}
\lefteqn{i \, \Delta^{R}_{F} (x-y) \quad := \quad 
  <0|T\,({\Psi}_R (x) \bar{\Psi}_L (y))|0> \quad =} \nonumber \\
 & \stackrel{!}{=} & 
i \int \! \frac{d^4p}{(2\pi )^4} \; e^{\displaystyle -ip (x-y)}
\frac{1}{\slsh{p} - M} =
i \int \! \frac{d^4p}{(2\pi )^4} \; e^{\displaystyle -ip (x-y)}
\frac{1}{p^2 - M^2} \; (\slsh{p} + M) \nonumber \\ 
 & & \nonumber \\
\lefteqn{i \, \Delta^{L}_{F} (y-x) \quad := \quad 
 - \; \gamma_0 \; {\left( <0|T\,({\Psi}_R (x) \bar{\Psi}_L (y))|0> \right) }^+
 \gamma_0 \quad =} \nonumber \\
 & \stackrel{!}{=} & 
i \int \! \frac{d^4p}{(2\pi )^4} \; e^{\displaystyle -ip (y-x)}
\frac{1}{\slsh{p} - M^\ast} =  
i \int \! \frac{d^4p}{(2\pi )^4} \; e^{\displaystyle -ip (y-x)}
\frac{1}{p^2 - M^{\ast\, 2}} \; (\slsh{p} + M^\ast) \qquad 
\end{eqnarray}
They obey the following equations:
\begin{eqnarray} (i \! \slsh{\partial}_x - M ) \; \Delta^{R}_{F} (x-y) \; = \; \delta^{\,
4}
(x-y) & , &
 \Delta^{R}_{F} (y-x) (- i \!\! \stackrel{\leftarrow}{\slsh{\partial}}_x - M )
 \;\; = \;
 \delta^{\, 4} (y-x)  
       \nonumber \\
(i \! \slsh{\partial}_x - M^\ast ) \; \Delta^{L}_{F} (x-y) \; = \;
      \delta^{\, 4} (x-y) 
 & , & 
 \Delta^{L}_{F} (y-x) (- i \!\! \stackrel{\leftarrow}{\slsh{\partial}}_x -
 M^\ast ) \; = \;
 \delta^{\, 4} (y-x) \qquad
\end{eqnarray}
\subsection{Effective model for a nucleon, a resonance and a
meson}
The model under consideration can now be extended by introduction of new degrees
of freedom, e.g. the nucleon field $N(x)$ (proton, neutron) and one meson 
$\phi_i (x)$ (internal index $i$), to obtain the following Lagrangian:
\begin{equation} {\cal L} (x) \quad = \quad {\cal L}^0_{N,N_\ast} (x) \quad
+ \quad {\cal L}^0_\Phi (x) \quad +
\quad {\cal L}^{\mbox{int}} (x) \end{equation}

\[ {\cal L}^0_{N,N_\ast} (x) = \left( \bar{N} (x), \bar{N}^R_\ast (x), \bar{N}^L_\ast
(x)\right) \; {\cal M} (N,N_\ast) \; 
\left( \begin{array}{l} N (x) \\ 
                       N^R_\ast (x) \\ 
                       N^L_\ast (x) \end{array} \right) \]
\begin{eqnarray} \lefteqn{{\cal L}^{\mbox{int}} (x)\quad =} \nonumber \\
 & = & - \;
\left( N^+ (x), N^{R\,+}_\ast (x), N^{L\,+}_\ast (x) \right) \!
\Bigg[ \Gamma^{\,i} (N,N_\ast) \, \Phi_i (x) + 
{\left( \Gamma^{\,i} (N,N_\ast) \right)}^+ \,\Phi^+_i (x)
\Bigg] \!\!
\left(\begin{array}{l} N (x) \\ 
                       N^R_\ast (x) \\ 
                       N^L_\ast (x)
\end{array}\right) \nonumber 
\end{eqnarray}
with the following $3\times 3$ matrices of Dirac-structures/operators:
\begin{eqnarray} {\cal M} (N,N_\ast) & := & \left(\begin{array}{ccc} {
\left( \frac{i}{2} \!\!\stackrel{\;\leftrightarrow}{\slsh{\partial}} - m
\right) } & 0 & 0 \\
 0 & 0 & {\alpha^\ast 
\left( \frac{i}{2} \!\!\stackrel{\;\leftrightarrow}{\slsh{\partial}} - M^\ast
\right) } \\
0 & {\alpha \left( \frac{i}{2} \!\!\stackrel{\;\leftrightarrow}{\slsh{\partial}} - M
\right) } & 0
\end{array} \right) \nonumber \\
 & & \nonumber \\
\Gamma^{\,i} (N,N_\ast) & := &
\left(\begin{array}{lcc} \frac{1}{2} \Gamma^{\,i}_{\Phi N\rightarrow N} & 0 & 0
\\
 \Gamma^{\,i}_{\Phi N\rightarrow N^R_\ast} & \frac{1}{2}\Gamma^{\,i}_{\Phi
 N^R_\ast\rightarrow N^R_\ast} & 0 \\
 \Gamma^{\,i}_{\Phi N\rightarrow N^L_\ast} & \Gamma^{\,i}_{\Phi
 N^R_\ast\rightarrow N^L_\ast} & 
 \frac{1}{2} \Gamma^{\,i}_{\Phi N^L_\ast\rightarrow N^L_\ast} 
\end{array}\right) 
\end{eqnarray}
$\Gamma^{\,i} (N,N_\ast)$ should be called ´´vertex matrix'' containing all vertex
structures between the fields considered. Summation over the internal indices $i$
of the meson field is required. The transition to the non-unitary
Wigner-Weisskopf approximation is performed by setting $N^{R\,+}_\ast
(x)=N^{L}_\ast (x)=0$.
\subsection{Implications to coupling constants}
As an example the non local interaction Lagrangian between the nucleon, the pion and the
Roper-resonance looks as follows:
\begin{eqnarray}
{\cal L}_{\pi NP_{11}} (x) 
& = & \frac{f_{\pi NP^L_{11}}}{m_{\pi}} \;
\left( \bar{N}^L_{P_{11}} (x) \; \gamma_\mu \gamma_5 \; \vec{\tau} \; N (x) 
\right) \; \cdot \; \partial^\mu {\vec{\Phi}}_{\pi} (x) \quad + \quad
 \nonumber \\
& + & 
\frac{f_{\pi NP^R_{11}}}{m_{\pi}} \;
\left( \bar{N}^R_{P_{11}} (x) \; \gamma_\mu \gamma_5 \; \vec{\tau} \; N (x) 
\right) \; \cdot \; \partial^\mu {\vec{\Phi}}_{\pi} (x) 
\quad + \quad
\mbox{h.c.} 
\end{eqnarray}
Assuming the pseudoscalar couplings $g_{\pi NP^L_{11}}$ and $g_{\pi NP^R_{11}}$
to be equal (arbitrary complex numbers), consistency within the model requires
the following relations between the pseudovector couplings:
\begin{eqnarray}
 & & \frac{f_{\pi NP^L_{11}}}{m_\pi} \quad = \quad  
\frac{g_{\pi NP^L_{11}}}{M_{P_{11}}+m_N}
 \quad , \quad  
 \frac{f_{\pi NP^R_{11}}}{m_\pi} \quad = \quad  
\frac{g_{\pi NP^R_{11}}}{M^\ast_{P_{11}}+m_N} \qquad , \nonumber \\
 & & \frac{f_{\pi NP^R_{11}}}{m_\pi} \quad = \quad  
  \frac{f_{\pi NP^L_{11}}}{m_\pi} \; \frac{M_{P_{11}}+m_N}{M^\ast_{P_{11}}+m_N}    
 \quad , \quad  
g_{\pi NP^R_{11}} \quad = \quad g_{\pi NP^L_{11}} \qquad \!\!\!\!  
\end{eqnarray}
Similar expressions hold for negative parity resonances, e.g. the 
$S_{11}(1535)$ resonance:
\begin{eqnarray}
 & & \frac{f_{\pi NS^L_{11}}}{m_\pi} \quad = \quad  
\frac{g_{\pi NS^L_{11}}}{M_{S_{11}}-m_N}
 \qquad , \qquad  
 \frac{f_{\pi NS^R_{11}}}{m_\pi} \quad = \quad  
\frac{g_{\pi NS^R_{11}}}{M^\ast_{S_{11}}-m_N} \qquad , \nonumber \\
 & & \nonumber \\
 & & \frac{f_{\pi NS^R_{11}}}{m_\pi} \quad = \quad  
  \frac{f_{\pi NS^L_{11}}}{m_\pi} \; \frac{M_{S_{11}}-m_N}{M^\ast_{S_{11}}-m_N}   
 \qquad , \qquad  
g_{\pi NS^R_{11}} \quad = \quad g_{\pi NS^L_{11}} 
\end{eqnarray}
Obviously the ``left'' and
``right'' pseudovector couplings differ by complex phases which are determined by
the resonance width. This property, 
{\em which has not properly been taken into account in literature upto
now}, should not be too surprising, as from renormalization theory it is
well known, that not only masses have to be renormalized, but also the coupling
constants. 
A very subtle point is still open for discussion:
as the interaction between nucleons and mesons can generate resonances as poles of
the S-matrix in the complex energy plane, one has to make clear -- {\em to avoid double
counting} --, in what way the effective resonant degrees of freedom in the model above
have to be interpreted.
\section{Final remarks}
As a result of the previous section it has been observed that the requirement of
unitarity leads to nontrivial constraints on self--energies and
vertex--functions which affect the interference between different subprocesses
in theoretical calculations and may be observed in experiments. For the
appropriate description of effective resonances the concept of the
Dirac--spinor has to be generalized and the number of independent effective
degrees of freedom per resonance has to be doubled. The extension of the toy
model to bosonic resonance fields is straight forward.
There are a various questions arising:\\
$\bullet\;$ Is there a way to find an effective intermediate energy field
theory of strong interaction including Chiral Perturbation Theory and baryon
effective degrees of freedom?\\
$\bullet\;$ Do baryonic resonance fields appear as effective degrees of freedom in an
effective field theory after integrating out the gluons from the generating
functional of QCD -- respecting the three and four gluon interaction terms ---
as mesons appear in Chiral Perturbation Theory after bosonising the quadratic
part of the general functional of QCD? How to avoid double counting problems
with respect to meson-nucleon-generated resonances? \\
$\bullet\;$ How is the toy model extended to momentum
dependent self--energies? \\
$\bullet\;$ Is it possible to extend the toy model to not so common
(non--linear) dispersion relations, which e.g.\ appear in thermal
field theories for time--like excitations \cite{bla1}, i.e.
\begin{equation} e^{-\,i\,\omega (\vec{p}\,)\,t} \rightarrow 
e^{-\,i\,\omega (\vec{p}\,)\,t} e^{-\gamma_t(\vec{p}\,)\,t} \quad \mbox{with}
\quad \gamma_t(\vec{p}\,) = \alpha T \ln(\omega_{p\ell} \, t) \quad , \quad
\omega_{p\ell} \sim g T
\end{equation}
$\bullet\;$ How to calculate vertex--functions on a microscopic basis (like
Sudakov did \cite{bro1}) in $NN$-physics? \\
$\bullet\;$ Is the QCD Sum Rule approach compatible with the idea of complex
vertex--functions? 

\end{document}